\def\PRL{{\it Phys. Rev. Lett.}\/ }
\def\JPCA{{\it J.Phys. C: Solid State Phy.}\/ }
\def\APL{{\it Appl. Phys. Lett.}\/ }
\def\JPN{{\it Jpn. J. Appl. Phys., Part 2}\/ }
\def\JMM{{\it J. Magn. Magn. Mater}\/ }
\def\JAP{{\it J. Appl. Phys.}\/ }
\def\PRB{{\it Phys. Rev. B}\/ }
\def\JPC{{\it J. Phys: Condens. Matter}\/ }
\def\JPD{{\it J. Phys.D: Appl. Phys.}\/ }
\def\bm{$\mu_B$\/ }
\def\no{\nonumber }
\def\eq{\ =\ }
\def\upa{\uparrow}
\def\upd{\downarrow}
\def\systema{Co$_2$Fe$_{0.4}$Cr$_{0.6}$Al\/ }
\def\systemd{Co$_2$Fe$_x$Cr$_{1-x}$Al\/ }
\def\systemb{Co$_2$CrAl\/ }
\def\systemc{Co$_2$FeAl\/ }

\def\etall{{\it et.al.}\/ }
\documentclass[singlespacing]{elsart}
\usepackage{graphics}
\usepackage{amssymb}
\begin{document}
\begin{frontmatter}
\title{The study of  electronic and magnetic properties of  
the partially disordered  pseudo-Heusler alloy \systema : an augmented space
approach}
\author[y1]{Monodeep Chakraborty}
\address[y1]{S.N. Bose National Centre for Basic Sciences, JD-III, Salt Lake City,Kolkata 700098, India}
\ead{monodeep@bose.res.in}
\author[x]{Atisdipankar Chakrabarti}
\address[x]{Ramakrishna Mission Vivekananda Centenary College, Rahara, West Bengal,
 India}
\ead{adc@bose.res.in}
\author[y2]{Abhijit Mookerjee}
\address[y2]{S.N. Bose National Centre for Basic Sciences, JD-III, Salt Lake City, Kolkata 700098, India}
\ead{abhijit@bose.res.in}

\begin{abstract}{In this communication we present a  study of the
electronic structure of partially disordered  
bulk  and (100) thin film  of quaternary pseudo-Heusler alloy 
\systema in the L2$_1$ phase
 using the Augmented Space recursion (ASR) in a scalar-relativistic tight binding linear muffin-tin 
orbitals (TB-LMTO) basis. 
 We study the orbital resolved
magnetic moment contributions of the  constituents of the
alloy. Our theoretical predictions match well with the
available experimental observations for the magnetic moments of
Fe and Co but they overestimate that of Cr.
For a (100) thin film,  layer as well as orbital resolved properties have been
studied.
}
\end{abstract}
\end{frontmatter}

\section{Introduction}   
Half metallic Heusler alloys (also known as half metallic ferromagnets, HMF) are those class of 
materials in which the majority spin band
is of metallic character, while the minority spin band is semiconducting, with a 
band gap at Fermi level ($E_F$). This remarkable property makes them potential
candidates for magnetic field sensors and spintronics devices. A large body
of literature have been devoted to experimental and theoretical studies of
such systems. The first compound to be identified as HMF by de Groot \etall \cite{r3_1}
was NiMnSb. Since then many different kinds of HMF have been reported, such as 
zinc-blende structured MnAs, CrAs, CrSb \cite{r3_2}-\cite{r3_3}, prevoskite structured
 La$_{0.7}$Sr$_{0.3}$MnO$_3$ \cite{r3_4}, 
rutile structured  CrO$_2$ \cite{r3_5} and  Co  based Co$_2$MnSn \cite{r3_6}. 

We focus our attention on 
the pseudo-Heusler alloys 
\systema. This is a pseudo HMF because unlike HMF it does not show
a true  gap at E$_F$ in the minority band. Still this is an important material
because of its ferromagnetic character at room temperature and at this particular
concentration it shows a high magneto-resistance ratio of up to $30\%$  at a
relatively low field of $0.1$T \cite{r3_7}-\cite{r3_8}. 
In addition it shows phase separation \cite{r3_9} and
tunnel magneto-resistance \cite{r3_10}-\cite{r3_11} at room temperature for the magnetic tunnel 
junction which utilizes \systema and Co$_2$MnAl as an electrode.

There have been several theoretical attempts in understanding the electronic
and magnetic behavior of Heusler alloys. Among them is the 
full-potential screened Korringa-Kohn-Rostocker(KKR) method in conjunction
with either coherent potential approximation or super cell construction
to account for the random distribution of  Cr  and  Fe  atoms. 
Results using fully relativistic version of the KKR-CPA formalism has been reported in
reference \cite{r3_7} for x = 0, 0.4 and 1.0 to study the magnetic nature of the alloy. 
They have studied the L2$_1$, B2 and A2  phases and their combinations 
separately. For the L2$_1$ phase and for x = 0.4 their
site projected magnetic moment for Co, Cr and Fe were 
0.94\bm, 1.42\bm and 2.92\bm respectively, whereas their experimental
values were 1.11\bm, 0.36\bm and 2.64\bm respectively. Reference \cite{anaton}
reported super cell based fully relativistic LMTO calculation of \systemd for x=0, 0.125, 0.25, 0.375, 0.5,0.625, 0.75 and 1. Their observation was, that the concentration and arrangement
of Fe played a decisive role in determining the magnetic properties of the alloy and its
constituents. Their theoretically calculated magnetic moments for Fe and Co were quite
near to their experimental values, but that for Cr was highly overestimated.
The large overestimation of the Cr moment is common to all the theoretical approaches. 
For all other phases their results also show
higher magnetic moment than the experimental observations.
Miura \etall \cite{mui} theoretically studied the \systema using KKR-CPA formalism.
Their findings indicate that spin polarization of \systemd at $E_{F}$, decreases
with increasing  Fe  concentration in both the L2$_1$ and B2 phases and
disorder effect plays a significant role on the spin polarization(at $E_{F}$)
at low Fe concentration. Elmers \etall \cite{elmer} used magnetic circular dichroism (MCD) 
and X-ray absorption spectroscopy (XAS) to study the \systema experimentally.
From these data they have also obtained site projected magnetic moments 
using magneto-optical sum rules and have compared  them  with band structure 
calculations. Recently there has been a spurt in experimental activity on 
thin films of \systemb and \systemc. Reference \cite{exp1}
reported magnetic properties of both $L2_{1}$ 
polycrystalline \systemb and epitaxial $L2_{1}$- structured
\systemc films on GaAs(001) substrate. Their observations showed
the existence of uniaxial magnetic anisotropy along the [1-10] for \systemc.
On the other hand \systemb showed an isotropic M-H loop.
In an another work Kelekar \etall \cite{exp2} reported a method of
development of single-phase epitaxial thin films for \systemd and 
observed large Hall resistivity in the range $(4-5)\times 10^{-8}$ $\Omega$  m
at 5 K for x=0.4.

\begin{figure}
\centering \resizebox{7.5cm}{7.5cm}{ \includegraphics{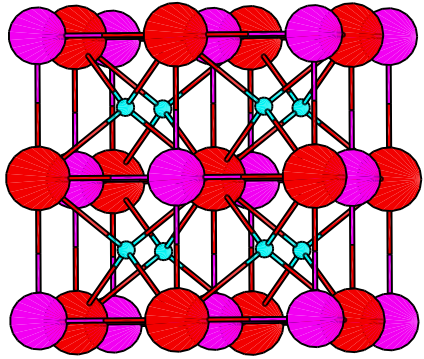}} \caption{The unit cell of a 
Heusler alloy X$_2$YZ where X and Y are transition metal atoms and Z a $sp$-metal atom. 
The medium sized spheres represent Al, large ones are either
Fe or Cr and he small spheres represent Co.} \label{fig1}
\end{figure}

In this communication we have investigated the magnetic and electronic properties
of the bulk and $100$ thin film of HMF \systema alloy, using TB-LMTO and ASR formalism.
Here we have taken $L2_{1}$ unit cell with $Fm\bar 3m$ symmetry, with experimental
lattice parameter $a\eq 5.727$ \AA \cite{r3_7}. The  Co  site is at $8c(\frac{1}{4},\frac{1}{4},\frac{1}{4})$
and the Al site is situated at $4a(0,0,0)$. The  Fe  and  Cr  occupy the $4b(\frac{1}{2},
\frac{1}{2},\frac{1}{2})$ site with probability $0.4$ and $0.6$, respectively. Figure \ref{fig1} shows the unit cell of a Heusler alloy.

This is an ideal case for for the study of partial disorder (PD) in the L2$_1$ phase. Here the 
$8c$ and $4a$ sites are occupied by the respective atoms with probability 1, and disorder
shows up in the $4b$ sites only. This consideration is of prime importance in order
to account for the occupancy of the sub lattice positions in L2$_1$ when we are away
from the stoichiometric case (\systemc or \systemb). 
To mimic a free standing Fe/Cr and Al terminated partially disordered
thin film of \systema, we have considered a nine layer thick 
films separated by seven layers of empty sphere.

\section {Methodology}
In this section we generalize the TB-LMTO-ASR formalism to many atoms per unit
cell, which is required to study systems having different disorder in different
sub lattices. 
Since the recursion method needs a localized, 
short-ranged basis for its operation, one can implement augmented space
recursion in the framework of the TB-LMTO formalism. 

The second order TB-LMTO Hamiltonian in the most localized representation
is given by, 

\begin{equation}
{\mathbf H}^{(2)}_{\sigma} = {E}_{\nu}^{\sigma} + h^{\sigma} -h^{\sigma}o^{\sigma}h^{\sigma}
\end{equation}
where, \\

\begin{eqnarray} h^{\sigma}  =  \sum_{RL\alpha} \left({ C}_{RL\alpha}^{\sigma} - 
 E_{\nu R L \alpha}^\sigma\right) \enskip {\mathbf P}_{RL\alpha} + 
\sum_{RL\alpha}\sum_{R'L'\alpha '} \Delta_{RL\alpha}^{\sigma\ 1/2}\ 
S_{RL\alpha, R'L'\alpha '}\ 
\Delta_{R'L'\alpha '}^{\sigma\ 1/2} \enskip {\mathbf T}_{RL\alpha, R'L'\alpha '}\nonumber\\
 \end{eqnarray}

\begin{itemize}
\item $R$ denotes a {\sl cell position} label associated with a TB-LMTO basis and $L$ = ($\ell m m_s$) is the
composite angular momentum index. 
\item $\sigma$ is the spin index and $\alpha$ denotes an atom in the $R$-th cell whose position is $R+\xi^\alpha$.
\item $C$, $o$ and $\Delta$ are potential parameters of the TB-LMTO method, 
these are diagonal matrices in the angular momentum indeces. 
Also $o^{-1}$ has the dimension of energy and is a measure of the energy window  around 
${\tilde E}$ in  which the approximate Hamiltonian ${\mathbf H}^{(2)}$ is reliable.
\item ${\mathbf P}_{RL\alpha}$ and ${\mathbf T}_{RL\alpha, R'L'\alpha '}$ are the projection and transfer
operators in Hilbert space $H$ spanned by tight-binding basis $\left\{\vert RL\alpha\sigma \rangle \right\}$.
\end{itemize}
To incorporate disorder in the system, we consider $C$, $o$ and $\Delta$
to be  random, while the structure matrix is non random. The justification
for  the non randomness of the structure matrix is that, we calculate properties of those
alloys, where the effect due to individual component size mismatch is negligible.
We introduce a site-occupation variable $n_R^\alpha$ which takes values 0 or 1
depending upon whether site $\alpha$ in the $R$-th cell is occupied by an A or a B atom. 
In the absence of short-range ordering, the probability density of these variables are
given by :

\[ \mathrm {Pr}(n_R^\alpha) \ =\ x^\alpha\ \delta(n_R^\alpha)\ +\ y^\alpha\ \delta(n_R^\alpha -1)\]

\noindent $x^\alpha$ and $y^\alpha$ are the concentrations of A and B components occupying the $\alpha$
labeled atom in the unit cell.

For {\sl partial disorder} this is a random variable whose probability density 
depends upon which sub lattice it belongs to, hence the
label $\alpha$ is associated with it. In terms of
$n_R^\alpha$ the random  site and angular momentum diagonal potential parameters take the following form:

\begin{eqnarray*}
 V_{RL\alpha}^{\sigma} & = & V_L^{A\sigma}\ n_R^{\alpha} + V_L^{B\sigma}\ \left( 1-n_R^\alpha \right) 
 =  V_L^{B\sigma} + \delta V_L^\sigma\ n_R^\alpha \nonumber\\
\delta V_L^\sigma & = & V_L^{A\sigma} - V_L^{B\sigma} \nonumber\\
\end{eqnarray*}
\noindent where, $V_{RL\alpha}^{\sigma}$ can be any one of $C_{RL\alpha}^{\sigma}$,
 $\Delta_{RL\alpha}^{\sigma\ 1/2}$, $o_{RL\alpha}^{\sigma}$ and 
$ E_{\nu RL\alpha}^{\sigma}$.

We now  obtain the Hamiltonian $ {\mathbf H}^{(2)}$ as a function of the random
occupation variables by inserting above expressions in $h$ and finally inserting $h$ in expression (1).
To set up the effective Hamiltonian from which we may obtain the configuration averaged
Green function, we follow the prescription of the Augmented space theorem \cite{AMook1}.
With each random variable $n_R^\alpha$ we associate an operator {\bf M}$_R^\alpha$ whose
spectral density is the probability density of $n_R$. The theorem tells us that :

\begin{equation} \ll G^\sigma(\{n_R^\alpha\},z)\gg\ =\ \langle \{\emptyset\}\vert \left(z\hat{\mathbf I} - 
\widehat{\mathbf H}^{(2)}_\sigma(\{\widehat{\mathbf M}_R^\alpha)\right)^{-1}\vert \{\emptyset\}\rangle\end{equation}

The augmented  Hamiltonian $\widehat{\mathbf H}^{(2)}$ 
is constructed by replacing the random site occupation variables $n_{R}^\alpha$ by
their corresponding operators $\widehat{\mathbf M}_R^\alpha$.  The eigenstates
$\vert 0^{R\alpha}\rangle$ and $\vert 1^{R\alpha}\rangle$ of $\widehat{\mathbf M}_R^\alpha$
 span the {\sl configuration space} $\phi^{R\alpha}$. The effective Hamiltonian $\widehat{\mathbf H}^{(2)}$ is an operator in the augmented space $\Psi = {\it H}\otimes \prod^\otimes\ \phi^{R\alpha}$, where
{\it H} is the space spanned by the TB-LMTO basis $\vert R\alpha\rangle$. 
space,  A representation of  $\widehat{\mathbf M}_{R}^\alpha$ is given by 

\begin{eqnarray}
 {\mathbf M}_R^\alpha \ &=&\  x^{\alpha} {\mathbf P}_{R\alpha}^{\uparrow} + y^{\alpha} {\mathbf P}_{R\alpha}^{\downarrow}+ \sqrt{x^\alpha y^\alpha}\ 
 {\mathbf T}_{R\alpha}^{\uparrow\downarrow}
\quad \epsilon \quad\phi^{R\alpha}\nonumber\\
\widehat{\mathbf M}_R^\alpha\ &=&\ {\mathbf I}\otimes\ldots {\mathbf M}_R^\alpha\otimes\ldots {\mathbf I}\otimes\ldots \quad\epsilon \quad\prod^\otimes\ \phi^{R\alpha}\ =\ \Phi
\end{eqnarray}

\noindent The representation is in  a basis $\vert\! \uparrow ^{R\alpha}\rangle = \sqrt{x^\alpha}\vert 0^{R\alpha}\rangle +
\sqrt{y^\alpha}\vert 1^{R\alpha}\rangle $  and $\vert\! \downarrow^{R\alpha} \rangle
 = \sqrt{y^\alpha}\vert 0\rangle -
\sqrt{x^\alpha}\vert 1^{R\alpha}\rangle $. 
 $P_{R\alpha}^\downarrow = \vert \downarrow^{R\alpha}\rangle\langle\downarrow^{R\alpha}\vert$ and
$T_{R\alpha}^{\uparrow\downarrow} = \vert\uparrow^{R\alpha}\rangle\langle\downarrow^{R\alpha}\vert +
\vert\downarrow^{R\alpha}\rangle\langle\uparrow^{R\alpha}\vert$ are the projection and transfer
operators in the configuration space $\Phi$
A general configuration state is of the type $\vert\upd\upa\upa\upd\ldots\rangle$. The sequence 
of sites $\{{\it C}\}$ where the configuration is $\upd$ uniquely describes a configuration. This is called the {\sl cardinality} sequence. 
The {\it average} configuration is one which has $\upa$ everywhere : or the null cardinality sequence $\vert\{\emptyset\}\rangle$.

A little algebra yields the following : If $V^\sigma_{RL\alpha}$ is a random potential
parameter, diagonal in
real and angular momentum space (as defined earlier) then we may define the following operators in configuration space :

\begin{eqnarray*}
\tilde{\bf A}(V^\sigma_{RL\alpha}) &= &\left( x^\alpha\ V^\sigma_{AL} + y^\alpha\ V^\sigma_{BL}\right)\ \tilde{\mathbf I}\otimes \ldots {\mathbf I} \otimes \ldots {\mathbf I}\ldots\\
\tilde{\bf B}(V^\sigma_{RL\alpha})& = &(y^\alpha -x^\alpha)\left(V^\sigma_{AL}-V^\sigma_{BL}\right)\ {\mathbf I}\otimes\ldots {\mathbf P}^\downarrow_{R\alpha}\otimes\ldots {\mathbf I}\otimes\ldots\\
\tilde{\bf F}(V^\sigma_{RL\alpha})& = &\sqrt{y^\alpha x^\alpha}\left(V^\sigma_{AL}-V^\sigma_{BL}\right)
\ {\mathbf I}\otimes\ldots {\mathbf T}^{\uparrow\downarrow}_{R\alpha}\otimes\ldots {\mathbf I}\otimes\ldots \\
\mathrm{and}\\
\tilde{\mathbf D}(V^\sigma_{R\alpha}) &= &\tilde{\mathbf A}(V^\sigma_{R\alpha})+
\tilde{\mathbf B}(V^\sigma_{R\alpha})+\tilde{\mathbf F}(V^\sigma_{R\alpha})
\end{eqnarray*}

The augmented space Hamiltonian then has the following compact form :

\begin{eqnarray*}
 \widehat{\mathbf H}^{(1)}_{\sigma} \ &=& \ \sum_{RL\alpha} \tilde{\mathbf D}(C^\sigma_{R\alpha}) \otimes {\mathbf P}_{RL\alpha}  \ldots\nonumber\\
& & + \sum_{RL\alpha}\sum_{R'L'\alpha '} \tilde{\mathbf D}(\Delta^{\sigma\ 1/2})_{RL\alpha} S_{RL\alpha,R'L'\alpha '} \tilde{\mathbf D}(\Delta^{\sigma\ 1/2}_{R'L'\alpha '})\otimes {\mathbf T}_{RL\alpha,R'L'\alpha'} \nonumber\\
\widehat{\mathbf h}^\sigma \ &=& \ \sum_{RL\alpha} \tilde{\mathbf D}(C^\sigma_{R\alpha}- E_{\nu RL\alpha}^\alpha) \otimes {\mathbf P}_{RL\alpha}  \ldots\nonumber\\
& &  + \sum_{RL\alpha}\sum_{R'L'\alpha '} \tilde{\mathbf D}(\Delta^{\sigma\ 1/2})_{RL\alpha} S_{RL\alpha,R'L'\alpha '} \tilde{\mathbf D}(\Delta^{\sigma\ 1/2}_{R'L'\alpha '}) \otimes {\mathbf T}_{RL\alpha,R'L'\alpha'} \nonumber\\
\widehat{\mathbf o}^\sigma \ &=& \ \sum_{RL\alpha} \tilde{\mathbf D}(o^\sigma_{R\alpha}) \otimes
{\mathbf P}_{RL\alpha} 
\end{eqnarray*}

Thus :

\begin{equation}
\widehat{\mathbf H}^{(2)}_\sigma \ =\ \widehat{\mathbf H}^{(1)}_\sigma - \widehat{\mathbf h}^\sigma\enskip
\widehat{\mathbf o}^\sigma\enskip  \widehat{\mathbf h}^\sigma 
\end{equation}

Using the augmented space theorem,  we can write the expression of 
configuration averaged Green function as, 

\[
 \ll G_{RL\alpha,RL\alpha}^{\sigma}(z) \gg  =  \langle 1 \vert\left(z\hat{\mathbf I} - \widehat{\mathbf H}^{(2)}_\sigma\right)^{-1} \vert 1 \rangle \]

\no where,

\[ \vert 1\rangle \ =\ \vert R \otimes L \otimes \alpha \otimes \{\emptyset\}\rangle \]

In order to obtain the Green function we shall use the recursion method of Haydock \etall \cite{hhk}.
This technique 
 transforms the sparse representation of the TB-LMTO Augmented space  
Hamiltonian  to a tridiagonal form.
This is done by constructing a  new orthonormal basis set $\vert n\}$ from the older one
$\vert n\rangle$ by the following three term recursion formula :

\begin{equation} 
\vert n+1\}=H\vert n\}+\alpha_n\vert n\}+\beta_{n-1}^2\vert n-1\}
\end{equation} 
with the initial choice $\vert 1\}$=$\vert 1\rangle$,
and $\beta_{0}^2=1$. 
 The coefficients $\alpha_n$ and $\beta_n$
are obtained by imposing the Otho-normalizability condition of the new basis set. They are given by :
\begin{equation}
 \frac{\{n\vert H\vert n\}}{\{n\vert n\}}=\alpha_n ;\hspace{0.6cm}
\frac{\{n\vert H\vert n-1\}}{\left[\{n\vert n\}\{n-1\vert n-1\}\right]^{1/2}}=\beta_{n-1}^2 ; \hspace{0.5cm} 
\{n\vert H\vert m\}=0\quad (m<n-1)
\end{equation}
Now the diagonal elements of the Green function can be calculated from the following expression:
\begin{equation} 
G_{RL\alpha,RL\alpha}^{\sigma}(z)= \frac{1}{\displaystyle z-\alpha_{1}-
                                \frac{\beta_{1}^2}{\displaystyle z-\alpha_2-
                                \frac{\beta_{2}^2}{\displaystyle z-\alpha_3-
\frac{\beta_3^2}{\displaystyle  \frac{\ddots}{\displaystyle z-\alpha_N-\beta_N^2 T(z)}}}}}
\end{equation} 

The above infinite continued fraction is terminated after a finite $n = N $(say) 
and the asymptotic part is replaced by a {\sl terminator} calculated from the first
N coefficients as suggested by  by Beer and Pettifor \cite{bp}.
 
The basis site, angular momentum projected  density of states (LDOS) is related to the configuration averaged Green function :

\begin{equation}
n^{\sigma}_{L\alpha}(E)=-\frac{1}{\pi}\ \Im m\  \lim_{\delta\rightarrow 0} \ll G^{\sigma}_{RL\alpha,RL\alpha}(E-\imath\delta^+)\gg 
\end{equation}

The Fermi energy $E_F$ is obtained from :

\[ \int_{-\infty}^{E_F}\ dE\ \sum_{L\alpha\sigma}\ n^{\sigma}_{L\alpha}(E)\ =\ \ll n\gg \]

\no where, $\ll n\gg$ is the average valence electrons per cell.

The basis site, angular momentum projected magnetic moment is obtained from :

\begin{equation} m_{L\alpha} \ =\ \int_{-\infty}^{E_F}\ dE \ \left(n^{\uparrow}_{L\alpha}(E)- n^{\downarrow}_{L\alpha}(E)\right) \end{equation} 
\begin{figure}
\centering
\resizebox{15cm}{6cm}{ \includegraphics{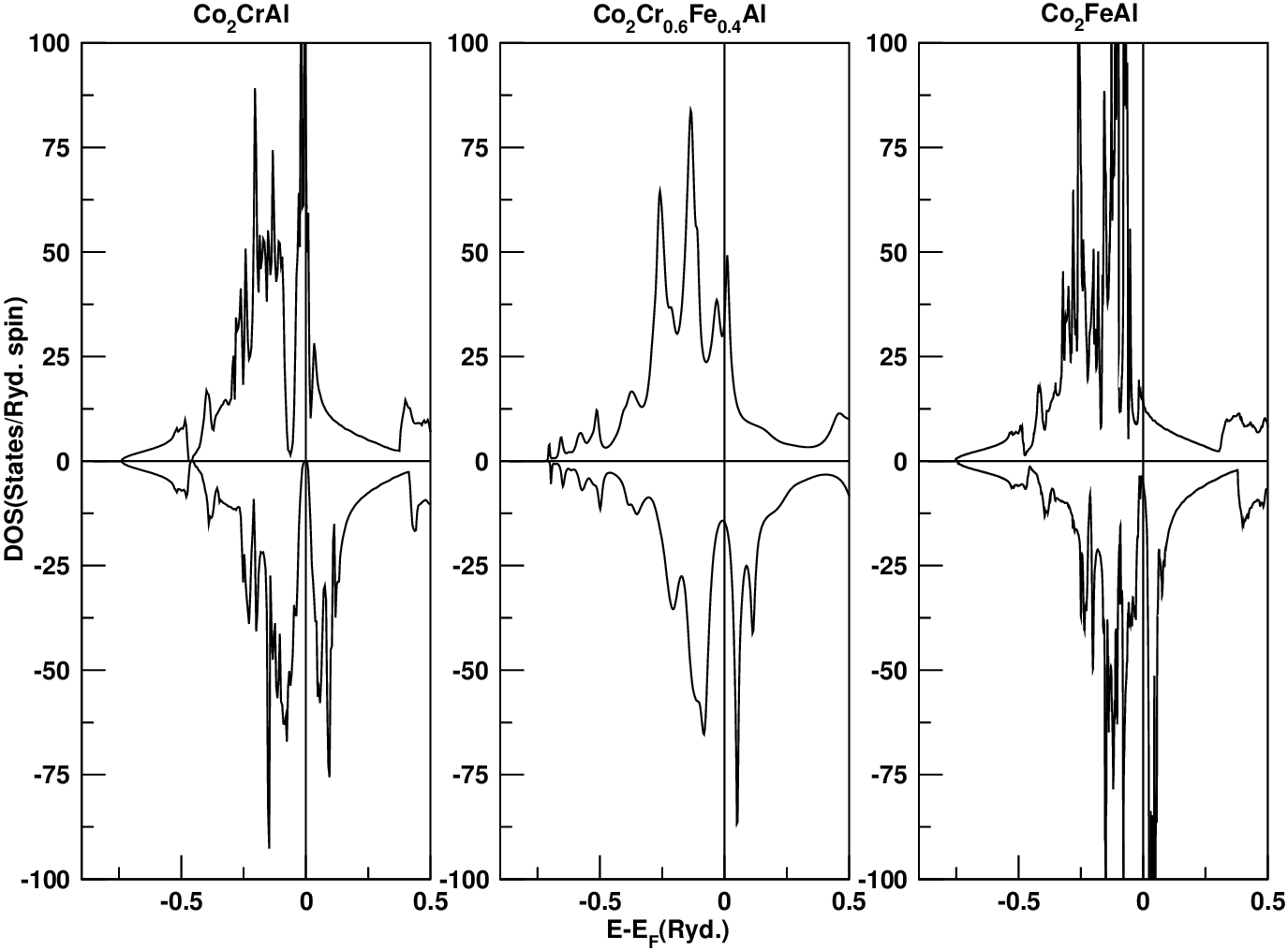}}
\caption{Total spin resolved DOS of \systemd ($x$ = 0,0.4,1) (center panel) compared with those of \systemb (left panel) and \systemc (right panel). Minority
states are shown on a negative scale}
\label{fh1}
\end{figure}

\section{Results for Bulk \systema}
\begin{figure}
\vskip 1cm
\centering
\resizebox{15cm}{12cm}{ \includegraphics{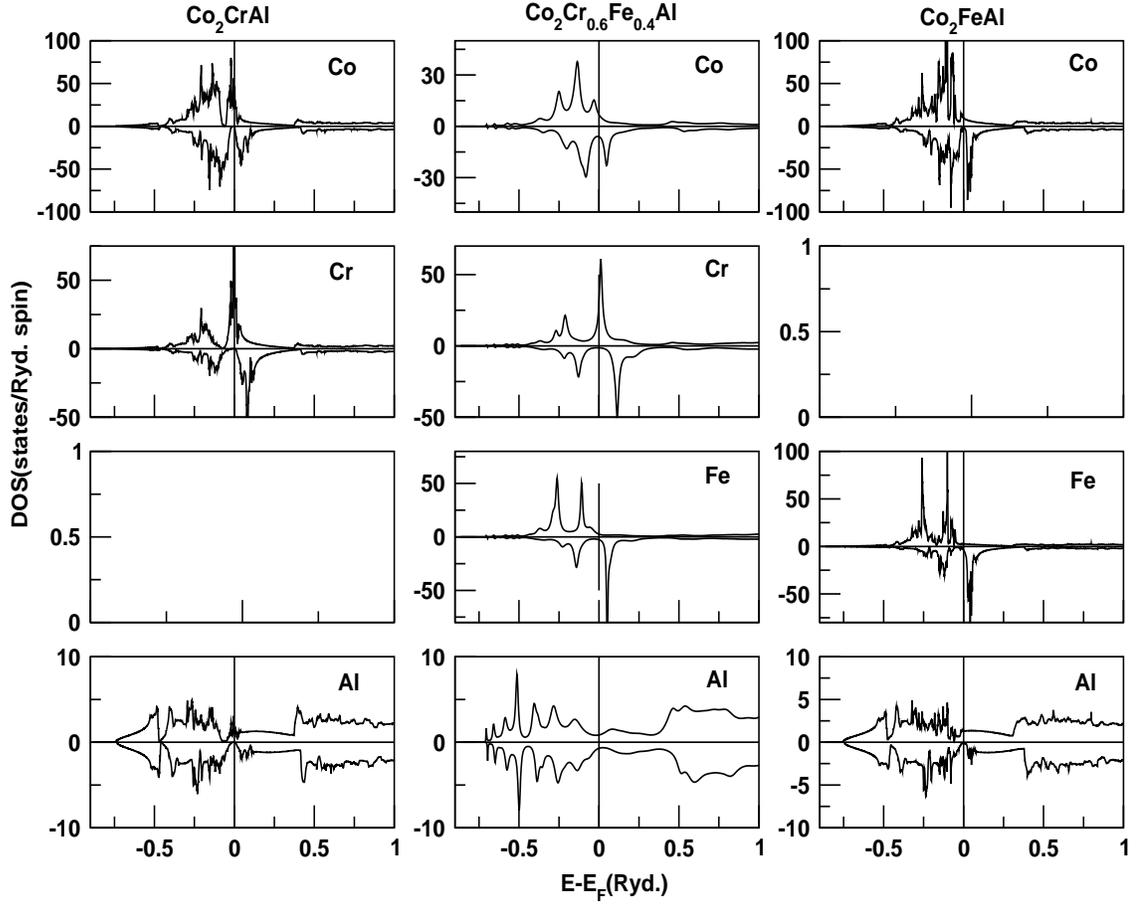}}
\caption{Site projected spin resolved DOS of \systemd (x = 0,0.4,1). Minority
states are shown on a negative scale}
\label{fh2}
\end{figure}

First, the electronic structure and magnetic properties of
pure compounds is compared with the L2$_1$ partially disordered
alloy. The spin resolved total DOS for the
three cases x = 0,0.4 and 1 are displayed in  figure \ref{fh1}.
It  shows that the Fermi level is situated in a valley in the minority spin density
of states. It is seen that for x = 0, spin minority band shows a gap which is a 
hall mark of HMF, but with increase in the  Fe  concentration
first the gap disappears and then for x = 1 it exhibits a 
sharp dip forming a pseudo gap. This can be understood from the fact, 
that with the introduction of 
 disorder in the $4b$ site, the Van Hove singularities are washed away
and consequently the band gap in the minority state is filled up. Yet,
vestiges of the kink singularities remain in the DOS as a signature of partial disorder.
The contribution from majority density of states at $E_F$
decreases with increase in  Fe  concentration. 
This decrease is monotonic in nature and maximum for \systemc.
A more detail study of the spin resolved site projected DOS (figure \ref{fh2})
 reveals that in \systemb the the bonding states in the minority band
is more Co like, where the anti-bonding state is dominated by  Cr .
The same feature has also been reported by earlier studies. For \systemc
the structures below the Fermi level are dominated by the Co majority band.
But above $E_{F}$ it has contribution from both Co and  Fe.
In the majority band we see that the two distinct peaks of  Fe 
reinforces the sharp peaks of Co. 
In the \systema the feature is little different. We see from figure \ref{fh2}
that both  Fe  and  Cr  almost retain their structures with respect
to parent compounds except the peak in majority band of  Cr  
at the $E_{F}$ has been shifted slightly towards right. This has major
significance in the magnetic moment of  Cr  which we will discuss later.

\begin{table}
\centering
\begin{tabular}{|l|c|c|c|c|c|c|c|c|c|c|}
%\hline
\multicolumn{11}{c}{{\bf \systema}} \\
\hline
 & \multicolumn{9}{c|}{Charge}& Mag.Mom \\
\hline
 & $s\uparrow$ & $p\uparrow$ &$d\uparrow$ &$s\downarrow$ &$p\downarrow$ &$d\downarrow$  & $\uparrow$ & $\downarrow$ & Tot. & ($\mu_B$) \\
\hline
 Co &0.346 &0.402 &4.38 &0.345 &0.429 &3.282 & 5.127 & 4.052 &9.179 & 1.075(1.11*)\\
\hline
 Cr  &0.294 & 0.370 & 2.544 & 0.299 & 0.397 &1.542 &3.209 &2.237 & 5.446 &0.972(0.36*)\\
\hline
 Fe  &0.348 & 0.437 & 4.591 & 0.345 & 0.459 &1.958 &5.38 &2.763 & 8.143 &2.617(2.64*) \\
\hline
Al &0.490 & 0.692 & 0.213 & 0.503 & 0.764 &0.244 &1.394 &1.511 & 2.905 &-0.117 \\
\hline
\multicolumn{11}{c}{ } \\
\multicolumn{11}{c}{{\bf \systemb}} \\
\hline
 & \multicolumn{9}{c|}{Charge}& Mag.Mom \\
\hline
 & $s\uparrow$ & $p\uparrow$ &$d\uparrow$ &$s\downarrow$ &$p\downarrow$ &$d\downarrow$  & $\uparrow$ & $\downarrow$ & Tot. & ($\mu_B$) \\
\hline
 Co &0.332 &0.402 &4.253 &0.320 &0.415 &3.492 & 4.987 & 4.227 &9.214 &0.760 (0.55*)\\
\hline
  Cr  &0.284 &0.369 &3.037 &0.284 &0.376 &1.482 & 3.69 & 2.142 & 5.832 &1.548(0.19*)\\
\hline
 Al &0.454 &0.675 &0.205 &0.466 &0.718 &0.221 & 1.334 & 1.405 &2.739 &-0.071\\

\hline
\multicolumn{11}{c}{ } \\
\multicolumn{11}{c}{{\bf \systemc}} \\
\hline
 & \multicolumn{9}{c|}{Charge}& Mag.Mom \\
\hline
 & $s\uparrow$ & $p\uparrow$ &$d\uparrow$ &$s\downarrow$ &$p\downarrow$ &$d\downarrow$  & $\uparrow$ & $\downarrow$ & Tot. & ($\mu_B$) \\
\hline
 Co &0.306 &0.377 &4.440 &0.312 &0.409 &3.283 & 5.123 & 4.004 &9.127 &1.119(1.57*)\\
\hline
  Fe  &0.333 &0.411 &4.634 &0.326 &0.432 &1.911 & 5.378 & 2.669 &8.047 &2.709(2.15*)\\
\hline
 Al &0.458 &0.646 &0.194 &0.464 &0.713 &0.222 & 1.298 & 1.399 &2.697& -0.101 \\
\hline
\end{tabular}
\label{tta}
\caption{The orbital resolved charges in  Fe  and  Cr  atomic spheres 
in Fe$_x$Cr$_{1-x}$ alloy. Tot. is site projected total charge. * are the corresponding
experimental values from  Wurhmehl \etall \cite{r3_7}.}
\end{table}

 The effect of alloying is very 
prominent in Al and Co. First we observe that the HMF like behavior
is lost due to filling up of gap at the $E_{F}$ in the
minority states in these two constituents.
Now we discuss the effect of charge transfer on the magnetism of \systema alloy.
Table-\ref{tta}  displays the element specific spin resolved charges and total
magnetic moment. For Co and  Fe  our calculated values are closest
to the experimental observations although the  Cr  moment is over-estimated.
This over-estimation of the Cr moment has been a recurrent feature of all earlier work. We shall examine whether this is due the accuracy of the electronic structure method used or the way we decide to treat the partial disorder in the system.

From Table-\ref{tta} we see that the 
magnetic moment of  Cr  in the \systema has decreased by an amount $0.576$ with respect to  \systemb.
This is due the fact that it has lost $0.392$ amount
of charge (with respect to pure compound) from  majority band. A more
careful study reveals that major portion of the charge is lost from
majority $d$ band. This is also evident in the  Cr  projected DOS shown
in figure \ref{fh2} where we see that the sharp peak at $E_{F}$ has been
pushed right wards from the Fermi level which accounts for the charge loss.
Since the DOS of  Fe  is almost same with respect to \systemc, it
also retains the value of its  magnetic moment. Al shows a small but 
 nonzero magnetic moment and it is oppositely polarized with respect to the 
other constituents in the \systema. 
%%%%%%%%%%%%%%%%%%%%%%%%%%%%%%%%%%%%%%%%%%%%%%%%%%%%%55
\begin{figure}
\centering
\resizebox{6cm}{5cm}{ \includegraphics{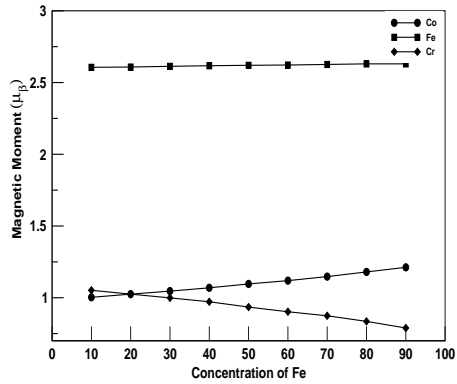}}
\caption{Variation of site-projected magnetic moments in \systema with
the variation in Fe/Cr concentration}
\label{magmom}
\end{figure}

We have also studied the variation of the site-projected magnetic moments as a
 function of the Fe-Cr disorder, which is shown in Figure \ref{magmom}.  The trends are very similar to those reported earlier (\cite{r3_7}). The overestimation of the Cr moment, in comparison with experiment, is consistent for all concentrations and of the same order of magnitude as earlier work using different theoretical techniques.  We shall comment on this in our concluding section.

\begin{figure*}
\vskip 1cm
\centering
\resizebox{10cm}{8cm}{ \includegraphics{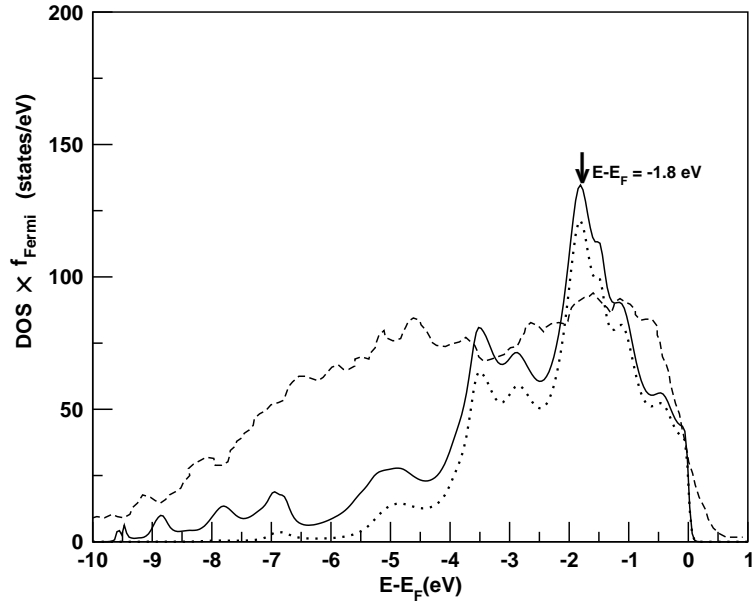}}
\caption{Comparison between the VB-XPS data of Wurhmehl \etall (\cite{r3_7}) (dashed lines) 
and the Fermi function convoluted total density of states (full lines) for \systema . The
dotted lines show the total $d$-projected DOS}
\label{fig11}
\end{figure*}

%%%%%%%%%%%%%%%%%%%%%%%%%%%%%%%%%%%%%%%%%%%%%%%%%%%%%55
It would be interesting to compare our density of states results
with experimental photoemmission spectra for \systema excited by
hard X-rays as reported by Wurmehl \cite{r3_7}. The experimental data
(Figure \ref{fig11}) indicate a  broad feature of width about 2 eV is seen just
below the Fermi energy. A second feature is seen between
4 eV and 7 eV below the Fermi energy.  The figure \ref{fig11} shows
the comparison between the VB-XPS data and the total density of states
convoluted with the Fermi function corresponding to the experimental temperature.   The density of states show two distinct features. First, the structure
due to the $d$-states around and just below the Fermi energy going down almost
5 eV below it. Below this are the features due to the $s$-$p$ states down to
10 eV below the Fermi energy. The figure shows that the total width of the structures due to the $d$-states has a width of about 7 eV, which is also the
conclusion from the high-resolution, high-energy spectra. The disagreement between the emission spectrum and the total DOS is qualitatively similar to the
work of Wurmehl \etall (\cite{r3_7}). The DOS shows a maximum at around 1.2 eV
below the Fermi energy, which is reproduced in the XPS. However, the
spectral feature lying from 4 to 7 eV below the Fermi energy is not reproduced at all by the local spin-density approximation based theoretical calculations. This disagreement is qualitatively similar to the earlier works references. Earlier works which ignored disorder suggested that disorder could be a factor, however, our work which includes disorder rules this out. We may speculate, with Wurmehl \etall \cite{r3_7} that this may
point to some deficiencies of the local spin-density approximation.
%%%%%%%%%%%%%%%%%%%%%%%%%%%%%%%%%%%%%%%%%%%%%%%%%%%%%55

\section{Thin film and surface of \systema}

%%%%%%%%%%%%%%%%%%%%%%%%%%%%%%%%%%%%%%%%%%%%%555

\begin{figure}
\centering
\resizebox{15cm}{12cm}{ \includegraphics{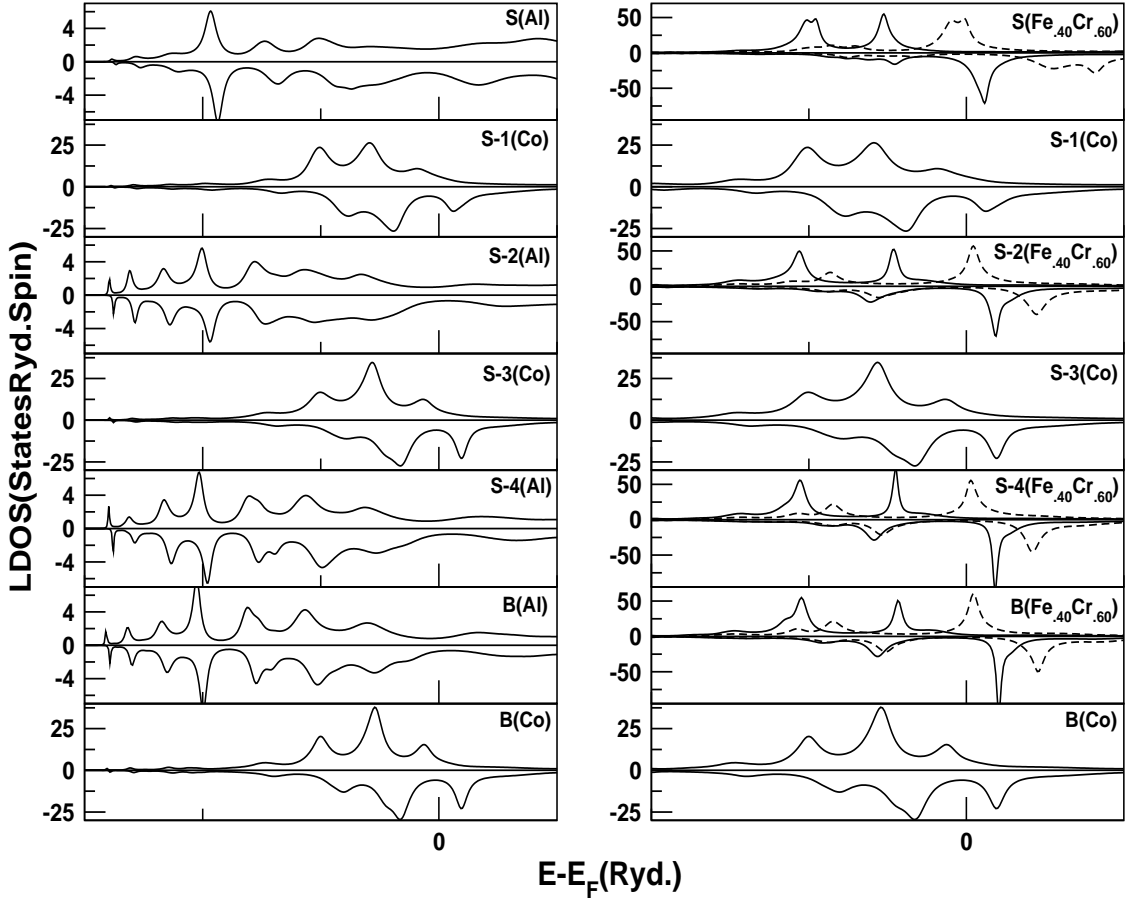}}
\caption{Site projected spin  and layer 
resolved DOS of \systema. Minority
states are shown on a negative scale}
\label{fh3}
\end{figure}

	The methodology we have used to study a (100) nine layered thin film
 of \systema is same as that of bulk. Here we have  carried out
 a TB-LMTO ASR calculation, to  obtain the layer, site as well as  orbital 
($\ell$-$m_l$-$m_s$) resolved density of states. The recursion method is ideal for
application to systems with surfaces. It was originally suggested by Haydock \etall
(\cite{hhk}) precisely for such a problem. 
With these results at our disposal we have performed a detailed analysis
of how  charge redistributes itself amongst the various orbitals as a result
of alloying in the Cr/Fe site and  due to the quasi two dimensional nature
of the system. The layer resolved DOS are shown in figure \ref{fh3} . All the three 
Surface states(Al, Fe  and  Cr ) exhibits characteristic surface narrowing
due to reduced coordination.  The majority surface state of  Cr  shows a
broadening at $E_F$ and is also slightly but importantly shifted towards left,
whereas the minority band has almost been pushed out of the $E_F$, and
 this is of prime importance, which results in a large increase in the  Cr  
magnetic
moment in surface. The majority surface state of  Fe  does not show much change
apart from the fact that the lowest lying peak has broadened out 
compared to bulk. The structure and height of the minority  surface 
 Fe  are less developed compared to its bulk counterpart below
$E_F$, whereas above the $E_F$ its structure though 
prominent, is broadened, left shifted and less sharp  than the bulk ones.
 The layer below the surface layer are occupied by Co and its 
relative peak width and height have changed with respect to bulk Co.
 As 
as we move down to the central layers, they start resembling the bulk
 DOS for
all types. This reflects a very important phenomenon of electronic structure,
that the Local density of states is exponentially insensitive to boundary
conditions, i.e., the Heine's black body theorem. Table-2  shows the
magnetic moments of  Cr , Fe  and Co in different sites
(layer projected)  of the
thin film and bulk . We can see the magnetic moment for the
 Cr  surface state is spectacularly enhanced whereas for the surface
  Fe , the enhancement is more conservative. The Co magnetic moment
below the surface is slightly suppressed compared to its bulk
value. From third layer onwards the magnetic moments  are more in
tune with their bulk counterparts, which is precisely reflected 
in the layer variation of the DOS.

                  Since both \systemc, \systemb and \systema have 
been considered in the $L2_1$
 phase 
, owing to the cubic symmetry   of the crystal structure in this phase, 
the five-fold degenerate $d$-orbitals  of the atomic case 
 is 
 broken into the three fold degenerate  $t_{2g}$ orbitals ($d_{xy}$,$d_{xz}$
 and $d_{yz}$)  and the two fold degenerate $e_g$ orbitals
($d_{x^2-y^2}$ and $d_{3z^2 -1}$).  
But for a (100) thin film  the z-direction is also rendered 
inequivalent with respect to x and y directions. This results in further
lifting of the degeneracies. Now the out of the five $d$-orbitals only two
states still retain their degeneracy ( $d_{xz}$ and $d_{yz}$) as a
signature of the remnant planar x-y symmetry of the smooth thin film.
 In a (100)  film the   three orbitals have their lobes
($d_{xz}$,$d_{yz}$and $d_{3z^2 -1}$) pointing towards the vacuum  and
other two ($d_{x^2-y^2}$ and $d_{xy}$)  have their probability distribution
 perpendicular to vacuum. 

\begin{figure}
\centering
\resizebox{15cm}{10cm}{ \includegraphics{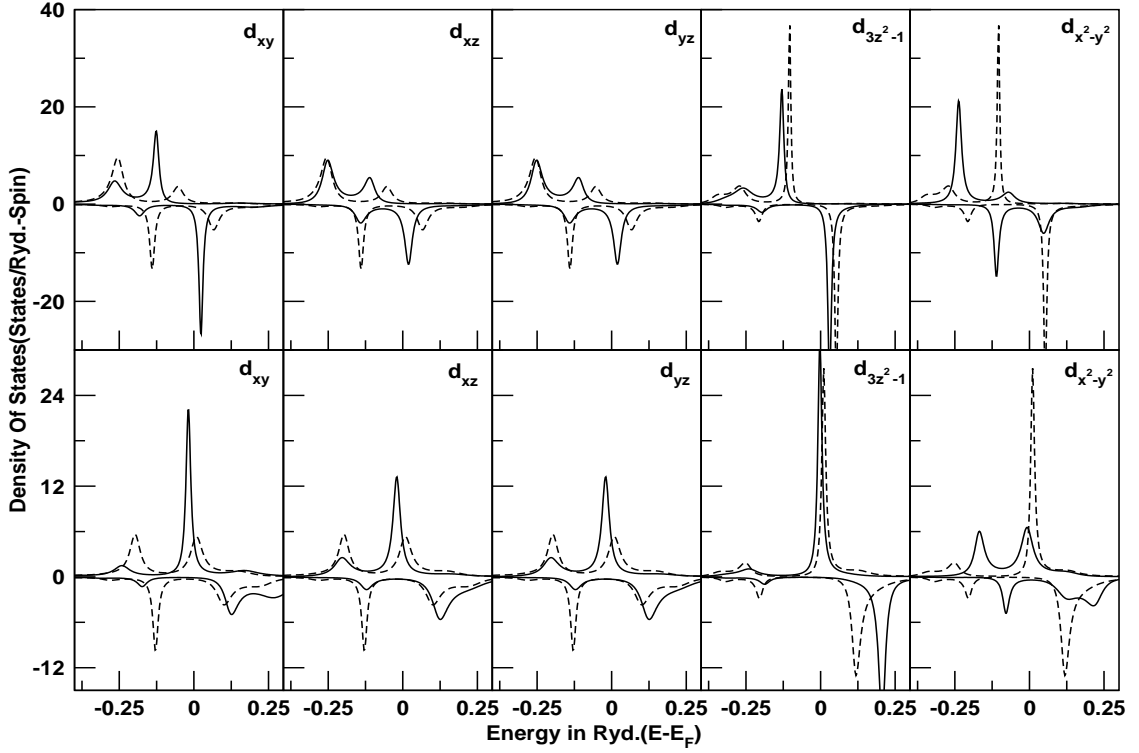}}
\caption{Spin resolved DOS of different d states of \systema. Minority
states are shown on a negative scale}
\label{fh4}
\end{figure}

Figure \ref{fh4} shows the spin resolved DOS for constituent
$d$-orbitals of surface  Fe (upper panel) and  Cr (lower panel) and 
are compared with their bulk 
counterparts. Table-3 shows the ($\ell$-$m_l$-$m_s$) resolved charges  and the resulting
magnetic moment for surface and bulk states of Fe and Cr. 

The  Cr   majority  $d_{xy}$,$d_{xz}$ and 
$d_{yz}$ surface states 
 show   reduction 
in  peak heights for the lower lying states within $E_F$ whereas  the peaks near $E_F$ shows a marked
 sharpening and
are importantly shifted to higher energies. 
 The same can be said about the
majority $d_{3z^2 -1}$ state.  
For the $d_{x^2-y^2}$ state, its 
peak near $E_F$ is also shifted toward the left, but not sharpened.  
 Below $E_F$ all the minority $d$-states of surface  Cr, with
 the exception 
of $d_{x^2-y^2}$, are less sharp   than corresponding
  bulk ones, 
whereas the structure of the DOS are more developed above $E_F$.
The $d_{x^2-y^2}$ state displays an opposite behavior.   Consequently,
 for  Cr  each of the
 majority $d$-orbitals
gain a sizable amount of charge compared to bulk, whereas for minority
$d$-orbitals each of them loses charge, except for a small gain  in
$d_{x^2 - y^2}$ state.  This  results in a large enhancement in magnetic moment
of  surface Cr . Since among all the $d$-orbitals minority state of $d_{xy}$
loses the maximum amount of charge, its enhancement is maximum.  The
surface enhancement of  Cr  though remarkable is not surprising, as even
for a anti-ferromagnetic thin film of pure  Cr , the surface  Cr  states 
exhibit a similar enhancement. 

%%%%%%%%%%%%%%%%%%%%%%%%%%%%%%%%%%%%%%%%
\begin{table}
\centering
\begin{tabular}{|c|c|c|c|}
\hline
 &  Cr  &  Fe  & Co  \\
\hline
 S &2.81 &2.80& \\
\hline
 S-1 & & & 0.89\\
\hline
 S-2 &0.87 & 2.59& \\
\hline
 S-3 & & & 1.067\\
\hline
 S-4 &0.90 & 2.57& \\
\hline
 B &0.97 & 2.62& 1.069\\
\hline
\end{tabular}
\label{ttc1}
\caption{The magnetic moment of  Fe  and  Cr  and Co for different
 layers of the thin film and Bulk.}
\end{table}
%%%%%%%%%%%%%%%%%%%%%%%%%%%%%%%%%%%%%%%%%%%%%%%%
                 
                Unlike   Cr   at the surface, where the individual
$d$-orbitals participate in surface enhancement,  the magnetic
moment of the  Fe  $d_{x^2-y^2}$ orbital is suppressed.
 The majority band of the $d_{x^2-y^2}$ state of a surface  Fe  has its 
peak near $E_F$ substantially reduced.
The main role in the magnetic suppression of  surface Fe $d_{x^2-y^2}$ state 
is played by its minority component. The density of states for minority  
 $d_{x^2-y^2}$ state is pushed below $E_F$ with
respect to the corresponding bulk state. This results 
 in  a gain in charge in the minority state and consequent reduction in the moment of Fe.

The Fe $d_{xz}$ and $d_{yz}$   at the surface share a similar fate by virtue of remnant planar symmetry. Their majority sates have their low 
lying peaks within $E_F$ almost at the same energy as that  of the 
corresponding bulk state, whereas the peaks near $E_F$, are shifted 
 to its left, resulting in small charge gain . Their minority band at the surface have their  peaks
near $E_F$ are sharpened as compared to the corresponding bulk state  
 resulting in small loss of charge compared to the  bulk. 
 This in turn results in
 small  moment enhancement for   $d_{xz}$ and $d_{yz}$. 
The  surface $d_{3z^2 -1}$
of   Fe  displays similar changes  as the  $d_{xz}$ and $d_{yz}$ states, when
compared to its bulk,
with the exception that its peak height near $E_F$ is less than that of its 
bulk. As a result its contribution to the surface enhancement is less than
  $d_{xz}$ and $d_{yz}$ states. 

The major contribution to the surface enhancement of  Fe  comes from  the $d_{xy}$ state. The reason for this is that its
minority band is almost  pushed out of $E_F$(i.e., to its right). In the majority band  the structure near $E_F$ is relatively  well developed(i.e., compared to its bulk state)  among all the $d$-states of surface  Fe atoms.  It is because of
 the $d_{xy}$ state,
that there is an enhancement in moment of surface  Fe,  
despite moment suppression in  the $d_{x^2 - y^2}$ state at the surface.
 
%%%%%%%%%%%%%%%%%%%%%%%%%%%%%%%%%%%%%%%%%%%%%555
\begin{table}
\centering
\begin{tabular}{|c|c|c|c|c|c|c|}
\multicolumn{7}{c}{\bf Surface} \\
\hline
$m_l$ & \multicolumn{2}{c|}{Charge}& Mom. & \multicolumn{2}{c|}{Charge}&Mom.\\
\hline
 &  Cr $\upa$ &  Cr $\upd$ & Cr &  Fe $\upa$ &  Fe $\upd$ & Fe \\
\hline
 $s$ &0.238&0.225 &0.014 &0.301 &0.285 &0.016 \\
\hline
 $p_{x}$ & 0.082&0.084 &-0.002 & 0.107&0.103 &0.004 \\
\hline
 $p_{y}$ & 0.082& 0.084&-0.002 &0.107 &0.103 &0.004 \\
\hline
 $p_{z}$ & 0.048&0.05 &-0.002 &0.059 &0.057 &0.002 \\
\hline
 $d_{xy}$ &0.720 &0.11 &0.61 &0.94 &0.285 &0.654 \\
\hline
 $d_{yz}$ &0.725&0.147 &0.578 &0.925 &0.464 &0.461 \\
\hline
 $d_{xz}$ &0.725&0.147 &0.578 &0.925 &0.464 &0.461 \\
\hline
 $d_{3z^{2}-1}$ &0.678 &0.084 &0.594 &0.972 &0.185 &0.787 \\
\hline
 $d_{x^{2}-y^{2}}$ &0.654 &0.211 &0.443 &0.913 &0.508 &0.405 \\
\hline
\multicolumn{7}{c}{ } \\
\multicolumn{7}{c}{{\bf Bulk}} \\
\hline
$m_l$ & \multicolumn{2}{c|}{Charge}& Mom. & \multicolumn{2}{c|}{Charge}&Mom.\\
\hline
 &  Cr $\upa$ &  Cr $\upd$ & Cr &  Fe $\upa$ &  Fe $\upd$ & Fe \\
\hline
 $s$ &0.294&0.299 &-0.005 &0.347 &0.345 &0.002 \\
\hline
 $p_{x}$ & 0.123&0.132 &-0.009 &0.145 &0.153 &-0.008 \\
\hline
 $p_{y}$ & 0.123&0.132 &-0.009 & 0.145&0.153 &-0.008 \\
\hline
 $p_{z}$ & 0.123&0.132 &-0.009 &0.145 &0.153 &-0.008 \\
\hline
 $d_{xy}$ & 0.572&0.392 &0.180 & 0.892&0.500 &0.392 \\
\hline
 $d_{yz}$ &0.572&0.392&0.180 &0.892 &0.500 &0.392 \\
\hline
 $d_{xz}$ & 0.572&0.392 &0.180 &0.892 &0.500 &0.392 \\
\hline
 $d_{3z^{2}-1}$ &0.414&0.182 &0.232 &0.959 &0.228 &0.731 \\
\hline
 $d_{x^{2}-y^{2}}$ &0.414&0.182 &0.232 &0.959 &0.228 &0.731 \\
\hline

\end{tabular}
\label{ttb1}
\caption{The orbital ($\ell $-$m_l$-$m_s$)  resolved  Fe  and  Cr 
projected charges
and corresponding magnetic moment for surface and bulk states}
\end{table}
%%%%%%%%%%%%%%%%%%%%%%%%%%%%%%%%%%%%%%%%%%%%%%%%

\section{Summary and Conclusion}

We have presented here a version of the TB-LMTO based augmented space recursion technique, modified to include systems with many atoms per unit cell and disorder
only in one specific lattice site in the basis. We have shown that the technique
is ideally suited to describe the pseudo-Heusler alloy system \systemd both 
in the bulk and at a surface. We examine the local, component projected
magnetic moments, which we believe are sensitively dependent on the accuracy with which we can describe the chemical environment of a component atom. Our predictions agree well with experiment and qualitatively with earlier work.
This is with the exception of the gross over-estimation of the Cr moment  for all disorder compositions. All earlier work also overestimated
the Cr moment quantitatively as much as ours. The question arises, do we understand why this is so ?

Miura \etal \cite{mui} studied the effect of disorder in the Al and Co sites on the behavior of \systemd. They found that disorder between Cr  and Al hardly changes matters, while the disorder between Cr and Co
 leads to a large decrease in the Cr moment. However Antonov \etal \cite{anaton} argue  that such disorder is highly unlikely energetically. That leaves us
 without a reasonable explanation for the large overestimation of Cr
 moment. It seems from our work that this discrepancy cannot lie at the door of the type of electronic structure method used (KKR or LMTO) or the method used to deal with the partial disorder (CPA, supercell or ASR).

We speculate that this overestimation may be either due to the use of the Density Functional approximation which cannot take into account correlation in the
localized $d$-states of the constituents properly. Alternatively, one
should examine the experimental data in some more detail to determine
 if sub-lattice disorder does exist because of the way the alloys have been prepared.                                                                                
For the atoms on a (100) surface of \systemd Cr shows a very large enhancement of its local magnetic moment. Fe also shows a moderate enhancement. We have analyzed the atom-spin projected densities of states to try to understand this phenomenon.

%%%%%%%%%%%%%%%%%%%%%%%%%%%%%%%%%%%%%%%%%%%%%%%%
\section*{Acknowledgments}
    We acknowledge with pleasure, the helpful discussions with Mr. Kartick
Tarafdar of S.N. Bose National Centre for Basic Sciences, Kolkata, regarding
energy independent ASR.  MC would like to acknowledge the
Department of Science and Technology, Govt. of India, for financial
support. ADC would like to thank 
Ramakrishna Mission Vivekananda Centenary College, Rahara, West Bengal, for encouraging the research project.
%%%%%%%%%%%%%%%%%%%%%%%%%%%%%%%%%%%%%%%%%%%%%%%%

\end{document}